\documentclass[%
 aip,
 amsmath,amssymb,
 reprint,%
]{revtex4-1}

\usepackage{ifsym}
\usepackage{wasysym}
\usepackage{amssymb}
\usepackage{amsmath}
\usepackage{dcolumn}
\usepackage{bm}

\usepackage{color}
\usepackage[pdftex]{graphicx}

\usepackage{mathtools}

\usepackage{makecell}

\usepackage{mathptmx}

\usepackage[dvipsnames]{xcolor}

\definecolor{darkblue}{rgb}{0,0,0.6}
\definecolor{darkred}{rgb}{0.6,0,0}
\usepackage[colorlinks=true,urlcolor=darkblue,citecolor=darkblue,linkcolor=darkred]{hyperref}

\usepackage{enumerate}
\usepackage{mathtools}
\usepackage{stmaryrd}

\usepackage{enumitem}

\usepackage{textcomp}
\usepackage{gensymb}

\newcommand{\ex}[1]{\mathrm{e}^{#1}}

\newcommand{\rr}[0]{\boldsymbol{r}}

\newcommand{\pp}[0]{\boldsymbol{p}}

\newcommand{\kB}[0]{k_{\mathrm{B}}}

\newcommand{\A}[0]{\text{A}}
\newcommand{\B}[0]{\text{B}}
\newcommand{\C}[0]{\text{C}}

\begin{document}

\title[]{Conditions for the propulsion of a colloid\\ surrounded by a mesoscale phase separation}
\author{Jeanne Decayeux}
%\affiliation{Sorbonne Universit\'e, CNRS, Laboratoire PHENIX, 4 place Jussieu, 75005 Paris, France}

%\affiliation{Sorbonne Universit\'e, CNRS, Laboratoire PHENIX, 4 place Jussieu, 75005 Paris, France}
\author{Marie Jardat}
%\affiliation{Sorbonne Universit\'e, CNRS, Laboratoire PHENIX, 4 place Jussieu, 75005 Paris, France}
\author{Pierre Illien}

\author{Vincent Dahirel}
\email{vincent.dahirel@sorbonne-universite.fr}
\affiliation{Sorbonne Universit\'e, CNRS, Laboratoire PHENIX (Physicochimie des Electrolytes et Nanosyst\`emes Interfaciaux), 4 place Jussieu, 75005 Paris, France}

\date{\today}

\begin{abstract}
   We study a two-dimensional model of an active isotropic colloid whose propulsion is linked to the interactions between solute particles of the bath. 
   The colloid catalyzes a chemical reaction in its vicinity, that yields a local phase separation of solute particles. The density fluctuations of solute particles result in the enhanced diffusion of the colloid. 
   Using numerical simulations, we thoroughly investigate the conditions under which activity occurs, and we establish a state diagram for the activity of the colloid as a function of the parameters of the model. We use the generated data to unravel a key observable that controls the existence and the intensity of activity: The filling fraction of the reaction area. 
   Remarkably, we finally show that propulsion also occurs in three-dimensional geometries, which confirms the interest of this mechanism for experimental applications.
    
\end{abstract}

\maketitle

\section{Introduction}

Many microscopic organisms are able to self-propel in order to perform various biological tasks \cite{Lauga2009}. Inspired by living systems, recent progress in physics and chemistry have resulted in the design of artificial self-propelled colloids \cite{Bechinger2016, Zottl2016a, Illien2017}. In this context, numerous theoretical or experimental studies have investigated \emph{anisotropic} colloids that generate gradients of solute concentration, temperature or electric potential responsible for propulsion \cite{Ruckner2007a,Golestanian2007}. This is exemplified by Janus colloids whose surface has asymmetric properties, for instance a catalytic and a noncatalytic side, which results in strong and persistent fluctuations in the density of the surrounding solute particles \cite{Ebbens2010,Samin2015,Wurger2015,Volpe2011,Buttinoni2012,Buttinoni2013,Oshanin2017,Jiang2010,Safaei2019,Popescu}.

Recently, the propulsion of \emph{isotropic} colloids has received a growing interest. It has been demonstrated that anisotropy is not necessary to achieve self-propulsion. In a system where a colloid  emits a solute (or catalyzes some reaction) isotropically around itself, fluctuations of solute (or reactant/product) density can arise in the vicinity of the colloid. These fluctuations can be amplified, thus breaking the symmetry of the bath surrounding the colloid. As a consequence, the colloid displays an intermediate anomalous diffusion and an enhanced diffusion on longer timescales \cite{Golestanian2009, Valeriani2013,Golestanian2019,DeBuyl2013a}. Spontaneous polarization can also come from the displacement of mobile catalysts attached to the surface of the colloid \cite{DeCorato2020}, the imbalance of surface tension in the vicinity of an interfacial swimmer \cite{Boniface2019}, or the nonlinear coupling between the solute density and the flow at the surface of the colloid \cite{Rednikov1994,Michelin2013,Michelin2014,Hu2019,Farutin2021}. This last mechanism has later been used experimentally to trigger the self-propulsion of large water droplets in an oil-surfactant medium \cite{Thutupalli2011,Izri2014,Herminghaus2014,Maass2016a,Illien2020,Izzet2020}.  

Despite this important body of literature on active colloids,  the solute-solute interactions are not taken into account in most theoretical studies addressing the propulsion of isotropic or anisotropic particles. Nevertheless, they can play a significant role, in particular in crowded environments, under confinement, or when they trigger a phase separation \cite{Zinn2020,Dattani2017,Semeraro2018}. Here, we study a mechanism in which the activity of 2D isotropic colloids arises from the interactions within the bath of solute particles. We consider an isotropic colloid, initially immersed in a bath of solute particles that interact with each other through purely repulsive interactions (these particles are denoted by $\A$). The colloid catalyzes a $\A\to\B$ reaction  in its vicinity that maintains the system in a non equilibrium situation. The $\B$ particles interact via a Lennard-Jones (LJ) potential. In this model, fluctuations of solute density close to the colloid  may be amplified when there is a phase separation in the LJ fluid. In bulk, the state of a LJ fluid depends on the strength of the attraction in the interaction potential (denoted by $\varepsilon$) and on the density of solute particles $\rho$. Interestingly, the diffusion of a 2D colloid in an infinite bath of attractive particles close to a phase transition has been shown to be a non monotonic function of the temperature of the fluid \cite{Torres-Carbajal2015,Torres-Carbajal2018}. In our model, a LJ fluid surrounds the colloid, but it is not a bulk fluid: The LJ particles are confined in a finite domain around the colloid. Therefore, we are modeling a phase separation at a mesoscopic scale, rather than a macroscopic phase transition. Such mesoscale phenomenon depends on the shape and size of the LJ domain. 
There exists sets of parameters for which the colloid self-propels when it is surrounded by this mesoscale phase separation \cite{Decayeux2021a}.

In this article, we investigate thoroughly the role of the different parameters of the model (density of solute particles, relative sizes of the colloid and solute particles, typical conversion rate from B to A), and elucidate the conditions under which self-propulsion occurs. In the case of 2D systems, we establish the state diagram for activity. %that were left aside so far:
Besides, we define an observable that controls the diffusion enhancement. It is based on the filling fraction of the reaction area.  This leads to the conclusion that the self propulsion mechanism relies mainly on the number of particles inside the reaction area. In addition to this extensive study of 2D systems, we show that, strikingly, the diffusion enhancement can be observed in 3D -- an important observation for potential experimental applications \cite{Dattani2017a}.

We describe the model in Section \ref{section-model}. In Section \ref{section-phase-diagram}, we give the state diagram that indicates the range of parameters for which activity occurs. We also
introduce and discuss the influence of the filling fraction of the reaction area. We sum up our results and analyse them with this new insight. In Sections \ref{section-size} and  \ref{section-TBA}, we investigate the influence of other parameters of the model: The characteristic time of the reverse reaction,  and the size difference between the solute and the colloid. We show that these parameters do not affect significantly the mechanism, which confirms its robustness. Finally, we examine a three-dimensional version of our system in Section \ref{section-3D}.

\section{Model and methods}
\label{section-model}

We study a two-dimensional system 
in a square box of length $l_{\mathrm{box}}$ with periodic boundary conditions. An isotropic colloid of diameter $\sigma_{\C}$ is surrounded by $N=500$ solute particles of diameter $\sigma_{\A}$. Both the colloid and the solute particles are embedded in an implicit solvent. The trajectories of particles are computed from Brownian dynamics simulations, where the overdamped Langevin equation is integrated using an Euler scheme \cite{ErmakJCP75,Frenkel}. The positions of each of the $N + 1$ particles in the system at time $t+\Delta t$ are deduced from the positions at time $t$ by
\begin{align*}
\rr_i(t+\Delta t)  = \rr_i(t) - & \frac{D_i}{k_{\rm B} T} \sum_{ i\neq j} \nabla U(|\rr_i  -\rr_j|) \Delta t \\
& + \sqrt{2 D_i \Delta t} \boldsymbol{\eta}_i 
\end{align*}
where $\rr_i$ is the position of particle $i$, $D_i$ its diffusion coefficient at infinite dilution (i.e. the bare diffusion coefficient), $U$ is the pair interaction potential, $k_{\rm B}$ the Boltzmann constant and $T$ the temperature. $\boldsymbol{\eta}_i$ is a random variable, chosen from a Gaussian distribution with zero mean, and variance equal to $1$.  
We use reduced units: Distances are measured in $\sigma_\A$, time in $D_\A / \sigma_\A^2$ with $D_\A$ the bare diffusion coefficient of particle A (i.e. $D_\A / \sigma_\A^2$ is the time needed by a solute particle to diffuse in an area of its diameter), and the energy is in $\kB T$. 

Initially, all the solute particles are of type $\A$ and interact with each other via a purely repulsive Weeks-Chandler-Anderson (WCA) potential:
\begin{equation}
    {U_{\text{WCA}} (r_{ij}) }= 
    \begin{cases}
     4\varepsilon' \left[ \left(\frac{d_{ij}}{r_{ij}}\right)^{12}-\left(\frac{d_{ij}}{r_{ij}}\right)^{6}  \right] & \text{if $r_{ij} \leq 2^{1/6}d_{ij}$,} \\
     0& \text{otherwise,}
    \end{cases}
\end{equation}
where $r_{ij}$ is the distance between $i$ and $j$, $d_{ij}=(\sigma_\A+\sigma_\C)/2$ if $i$ or $j$ is the colloid, and $d_{ij}=\sigma_\A$ otherwise. 

The colloid C triggers a local phase separation in its vicinity by  catalyzing isotropically the reaction $\A+\C \to \B+\C$ in a reaction area of radius $r_{\mathrm{cut}}$ around itself. The solute particles of type B interact with each other via a short-ranged attractive Lennard-Jones (LJ) potential:
\begin{equation}
   {U_{\text{LJ}} (r_{ij}) }= 4\varepsilon \left[ \left(\frac{\sigma_\A}{r_{ij}}\right)^{12}-\left(\frac{\sigma_\A}{r_{ij}}\right)^{6}  \right] ,
\end{equation}
 for which we set a cut off for $r_{ij} \geq 2.5 d_{ij}$ to reduce computational costs, as the long-range effects are  negligible here.
 We choose $\varepsilon'=10$, and $\varepsilon$ varies.

 Outside the reaction area, the reverse reaction $\B +\C \to \A +\C$ takes place. This allows us to model a system where there is a constant supply of $\A$ solute particles that play the role of a \emph{fuel} for propulsion. The $\A \to \B$ and $\B \to \A$ reactions, which take place respectively inside and outside the reaction area, occur at exponentially distributed random times of respective averages $\tau_{\A \B}$ and $\tau_{\B \A}$. The average reaction times are taken identical and equal to $0.1$, except in Sec. \ref{section-TBA} where the influence of the value of $\tau_{\B \A}$ is studied. The reactions then occur  very fast compared to the other timescales of the problem.

Under these conditions, a steady state is reached where a Lennard-Jones fluid forms  in the reaction area. Outside the reaction area, there is a suspension of $\A$ particles, which interact with each other via a purely repulsive potential. The structure of the Lennard-Jones fluid in the reaction area depends on the average solute density $\rho=N/l_{\rm box}^2$, on the intensity of the attraction $\varepsilon$, and on the size of the reaction area, controlled by $r_{\mathrm{cut}}$. We have shown  that for some sets of parameters, namely $\rho=0.1$, and $\varepsilon$ varying from $2\kB T$ to  $3\kB T$, self-propulsion is observed \cite{Decayeux2021a}. In this case, the long-time diffusion coefficient of the colloid, denoted by  $D_{\rm eff}$ 
and defined by 
\begin{equation}
    D_{\mathrm{eff}} \equiv \lim_{t\to\infty} \frac{1}{4t} \langle \left[  \rr_\text{C}(t)- \rr_\text{C}(0) \right]^2\rangle,
    \label{def_Deff}
\end{equation}
where $ \rr_\text{C}(t)$ the position of the colloid at time $t$, is larger than its value without reaction in the same solute bath $D_{\mathrm{noreac}}$. Indeed, in the reaction area, the Lennard-Jones fluid demixes, which creates strong density fluctuations in the vicinity of the colloid. Solute particles polarize and form one or several droplets, which push the colloid in the opposite direction. If the orientation of droplets  persists for long enough, the motion of the colloid is transiently ballistic,  and becomes diffusive at long times, with $D_{\mathrm{eff}}\gg D_{\mathrm{noreac}}$.

In our previous paper \cite{Decayeux2021a}, we have shown that the ballistic motion of the colloid observed at intermediate timescales was related to the persistent orientation of a polarization vector $\pp$, that represents the polarization of solute particles around the colloid and is defined by
\begin{equation}
    \pp=\sum_{i\in\mathcal{P}} [\rr_i(t) - \rr_\text{C}(t)],
    \label{def-p}
\end{equation}
where $\rr_i(t)$ is the position vector of solute $i$, $\mathcal{P}$ is the circular area around the colloid, 
where solute particles may interact directly with the colloid (we have chosen the radius of this area to be $ (3\sigma_{\A}+\sigma_{\C})/2$). In what follows, we also compute the auto-correlation function of this polarization vector  $\langle \pp(0)\cdot\pp(t) \rangle $. An example of this correlation function as a function of time  can be found in Section \ref{section-TBA} (Fig.~\ref{taup_tauBA}) and will be commented later. In any case, this auto-correlation function is found to decay as a power law at short times, and  exponentially at  sufficiently large times. We define the characteristic time $\tau_p$ as the orientation persistence time. We extract $\tau_p$ from an exponential fit of the auto-correlation function $\langle \pp(0)\cdot\pp(t) \rangle  \propto \ex{-t/\tau_p} $ at long times.

The simulation procedure is the following. First, simulations of the system  without reaction are performed until the equilibrium is reached. 
Second, simulations  with reaction are run starting from independent configurations taken from the equilibrium trajectories without reaction. Then, once a steady state is reached in the situation with reaction, the effective diffusion of the colloid, $D_{\mathrm{eff}}$, and the auto-correlation of the polarization vector, $\langle \pp(0)\cdot\pp(t) \rangle$, are computed. The results are averaged over all the independent realisations (about $500$ realisations in any case). We define an error $\delta D$ on the estimation of the long-time diffusion coefficient (resp. the diffusion coefficient without reaction) by computing three values of $D_{\mathrm{eff}}$ (resp. $D_{\text{noreac}}$) for a smaller number of realisations and by taking the standard deviation of these three values. We assume that activity occurs if $D_{\mathrm{eff}} - \delta D > D_{\mathrm{noreac}} + \delta D $.

\section{Towards an observable to describe the influence of solute on colloid activity}
\label{section-phase-diagram}

\subsection{Local density around the colloid}

As we will discuss in the next sections, the presence of activity and the value of $D_{\mathrm{eff}}/D_{\mathrm{noreac}}$ are strongly related to the fraction of the reaction area that is filled with solute particles. In this section, we introduce a parameter that enables a better understanding of the propulsion mechanism: The number of particles inside the reaction area, $N_{\mathrm{shell}}$. %This quantity depends on time, and reaches an almost constant value at stationary state.
In order to compare the different systems with each other, $N_{\mathrm{shell}}$ is normalized by the maximum number of solute particles that would fill the reaction area, denoted by $N_{\mathrm{max}}$.  $N_{\mathrm{max}}$  depends on the reaction radius $r_{\mathrm{cut}}$, and on the maximum surface packing fraction
$\phi_{\rm max}$ through 
\begin{equation}
N_{\mathrm{max}}=\phi_{\rm max}\frac{r_{\mathrm{cut}}^2-(\sigma_{\mathrm{C}}/2)^2}{(\sigma_{\mathrm{A}}/2)^2},
\label{Nmax}
\end{equation}
{where $\phi_{\rm max}=\pi\sqrt{3}/6\simeq 0.91$ is the maximal packing fraction for hard disks placed on a 2D hexagonal lattice.}

\begin{figure}%[!h]
\includegraphics[width=\columnwidth]{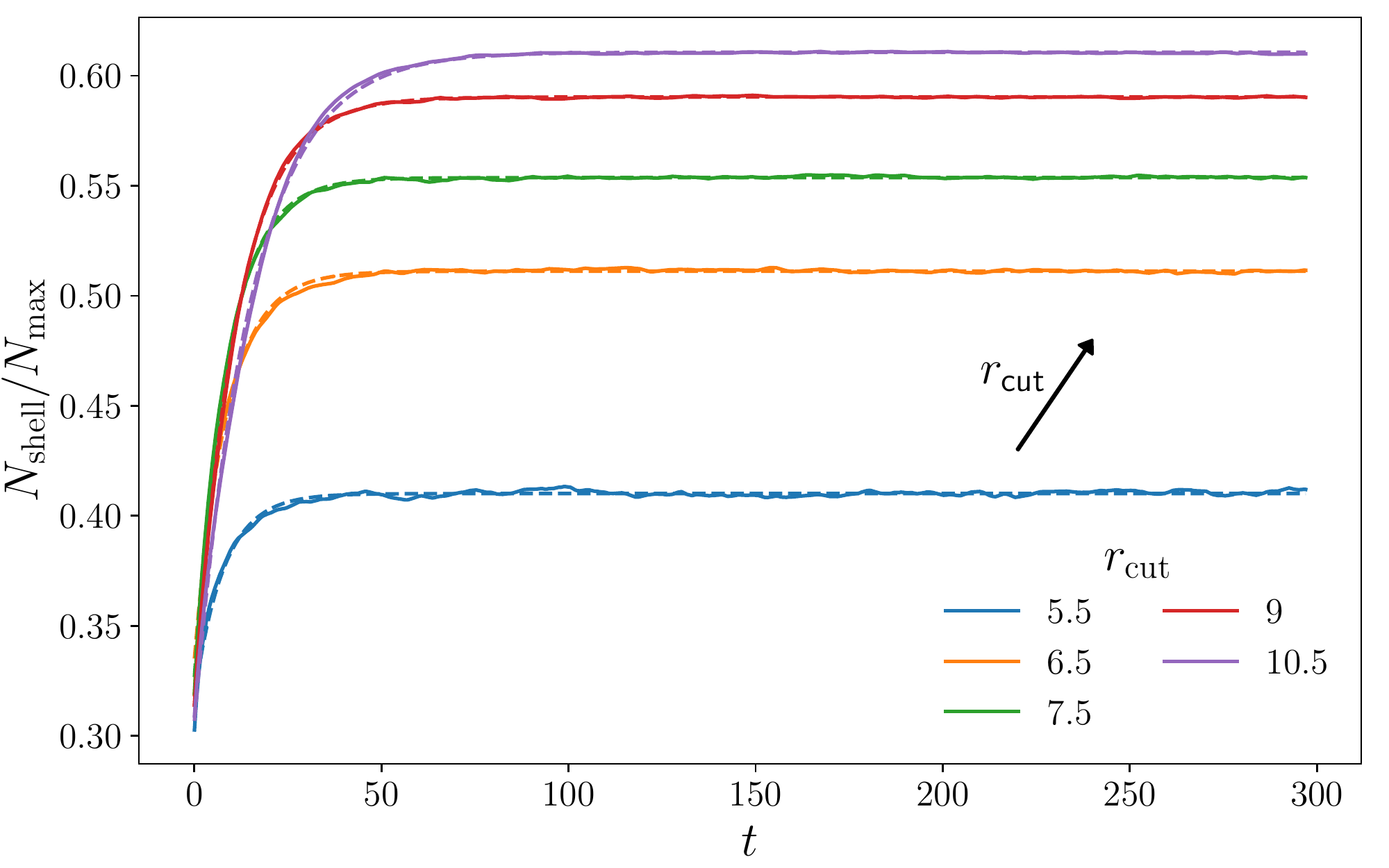}
\caption{Number of particles in the reaction area of radius $r_{\rm cut}$, $N_{\mathrm{shell}}$,  divided by $N_{\mathrm{max}}$, the maximum number of particles that would fill the reaction area (defined in Eq. \eqref{Nmax}), as a function of time. The plots correspond to different values of $r_{\rm cut}$, for $\rho=0.3$ and  $\varepsilon=3$. The dashed lines are exponential fits of the simulation data.  Note that, on this plot, the initial time $t=0$ represents the very beginning of the simulation with reaction. 
}
\label{n_shell}
\end{figure}

\begin{figure}
\includegraphics[width=\columnwidth]{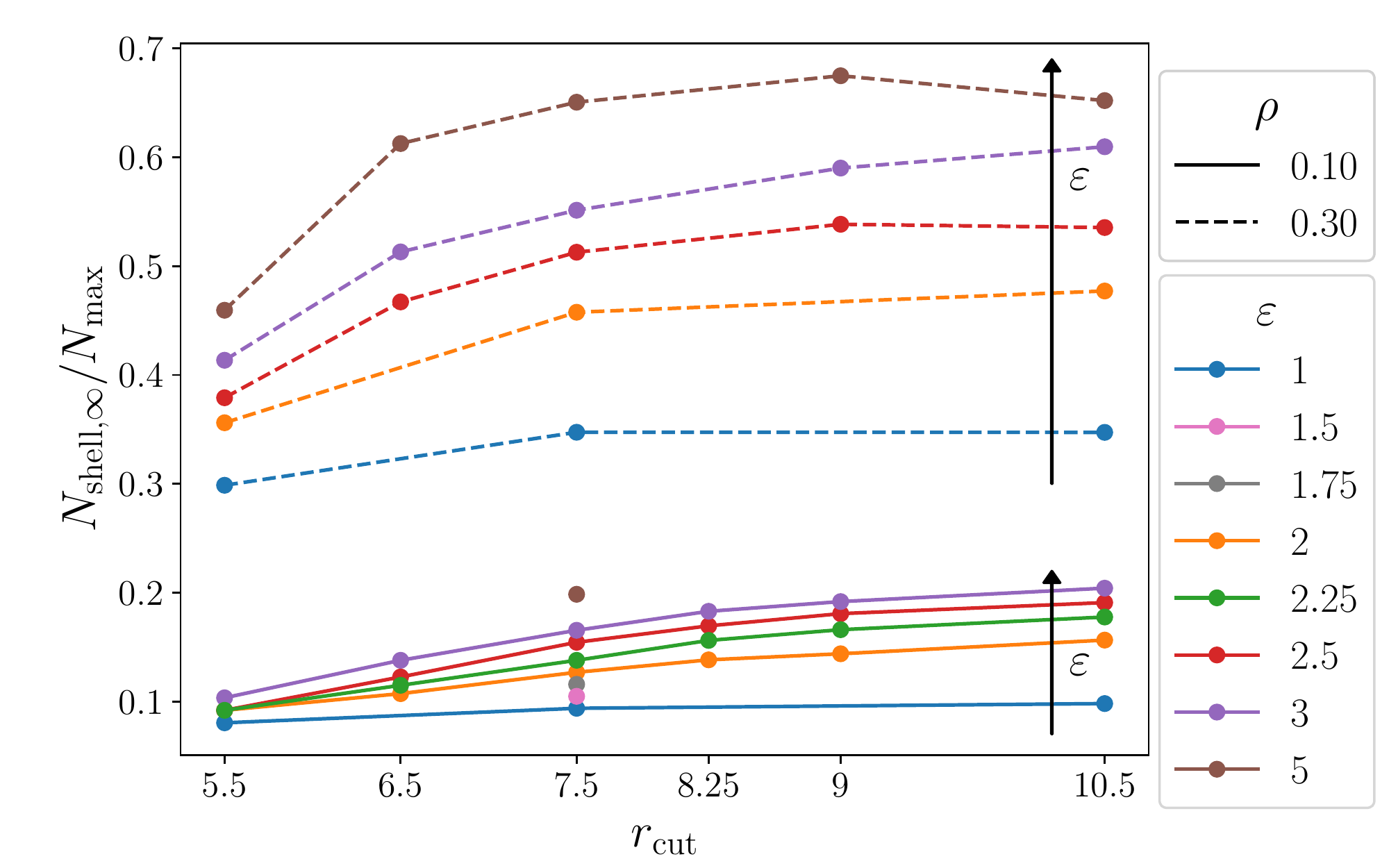}
\caption{Stationary value of the number of particles in the reaction area, $N_{\mathrm{shell},\infty}$,  divided by $N_{\mathrm{max}}$, the maximum number of particles that would fill the reaction area, as a function of $r_{\rm cut}$. The curves correspond to different values of $\varepsilon$ (different colors), for $\rho=0.1$ plain lines and $\rho=0.3$ dashed lines.
}
\label{n_shell_inf}
\end{figure}

The kinetics of the filling of the reaction area is shown on Fig.~\ref{n_shell} for fixed values of the density ($\rho=0.3$) and of the  interaction strength between LJ particles ($\varepsilon=3$), and for different values of $r_{\mathrm{cut}}$. Starting from an initial equilibrium situation where the colloid is embedded in a bath of A solute particles without any reaction, the number of particles inside the reaction area increases with time when the reaction occurs, due to the LJ attraction between B solute particles.  
At steady state, the colloid accumulates new particles in its vicinity as it moves, and loses as much as it displaces its reaction area with itself. 
$N_{\mathrm{shell}}$/$N_{\mathrm{max}}$ reaches a stationary value $N_{\mathrm{shell},\infty}/N_{\mathrm{max}}$ with a characteristic time $\tau_{\mathrm{N}}$.  This process can be modelled by the following exponential behavior:
\begin{equation}
\frac{N_{\mathrm{shell}}(t)}{N_{\rm max}}= \frac{N_{\mathrm{shell},\infty}}{N_{\rm max}} - \frac{N_{\mathrm{shell},\infty}-N_{\mathrm{0}}}{N_{\rm max}} \exp(-t/\tau_{\mathrm{N}})
\label{eq_n_shell}
\end{equation}
where $N_0=N(0)$ is the initial number of particles in the reaction area. The corresponding exponential fits are
represented by the dashed lines on Fig.~\ref{n_shell} and are in good agreement with computed data. The stationary value $N_{\mathrm{shell},\infty}/N_{\mathrm{max}}$ is an increasing function of $\varepsilon$ and increases for small sizes of the reaction area $r_{\rm cut}$ before reaching a plateau (see Fig.~\ref{n_shell_inf}). In what follows,  time dependent quantities are all computed at the steady state. Consequently, the simulation time is shifted so that the new initial time $t=0$ corresponds to about $3 \tau_{\mathrm{N}}$.

\subsection{State diagram for activity}

\begin{figure}[!h]
\includegraphics[width=\columnwidth]{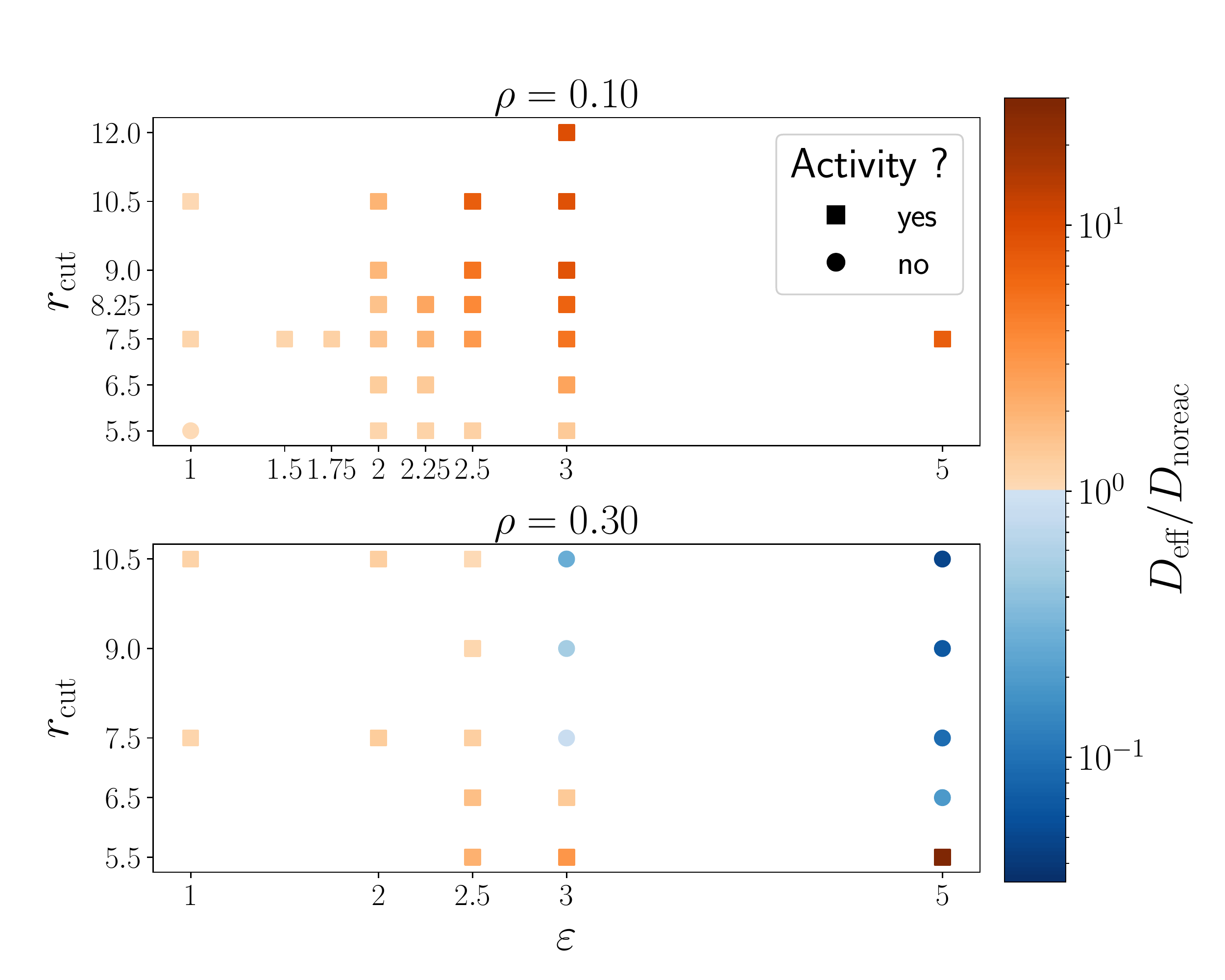}
\caption{State diagram of the studied systems at $\rho=0.1$ (top) and $\rho=0.3$ (bottom). Colors indicate the value of $D_{\mathrm{eff}}/D_{\mathrm{noreac}}$, where $D_{\mathrm{eff}}$ is the long-time diffusion coefficient of the colloid (defined in Eq. \eqref{def_Deff}) and $D_{\mathrm{noreac}}$ its value in the absence of reaction, from lower values (blue) to higher ones (dark orange). Square symbols indicate systems where activity was reported, circles systems without activity. Colloid displays activity if $D_{\mathrm{eff}} - \delta D > D_{\mathrm{noreac}} + \delta D $.}
\label{map}
\end{figure}

The influence of the parameters $\varepsilon$ and $r_{\mathrm{cut}}$ is summed up on the diagrams on Fig.~\ref{map} for two solute densities: $\rho=0.1$ (top) and $\rho=0.3$ (bottom). The color represents the relative value of the effective diffusion coefficient compared to its equilibrium value $D_{\mathrm{noreac}}$: Colors are ranging from blue (when $D_\text{eff}<D_\text{noreac}$) to dark orange (when $D_\text{eff}>D_\text{noreac}$). Symbols indicate if activity is observed (squares where activity occurs, circles if not).

As it appears on Fig.~\ref{map}-top, at a relatively low solute density ($\rho=0.1$, with $l_\text{box}=70$), activity occurs for all the parameters tested, and is increased when $\varepsilon$ or $r_{\mathrm{cut}}$ increases. 
The results obtained at a higher density of solute,  $\rho=0.3$, are displayed on Fig.~\ref{map}-bottom. To increase the density, the size of the simulation box was decreased, keeping the same amount of solute particles ($l_\text{box}=40$). Again, simulations with different values of the parameters $\varepsilon$ and $r_{\mathrm{cut}}$ were done.
We observe that the propulsion is more difficult to achieve for this higher solute density, as the range of parameters where $D_\text{eff}$ is significantly higher than $D_{\mathrm{noreac}}$ is more restricted. For example, activity occurs for all values of $r_{\rm cut}$ investigated here at $\varepsilon=2.5$, but disappears for large reaction areas ($r_{\rm cut}\ge 7.5$) at  $\varepsilon=3$. Indeed, at  a solute density $\rho=0.3$, the reaction area is more likely to be densely filled with $\B$ particles  than when $\rho=0.1$. The dense fluid that occupies the reaction area tends to hinder the motion of the colloid.

\subsection{A normalized local density as a key parameter to predict activity}

\begin{figure}
\includegraphics[width=\columnwidth]{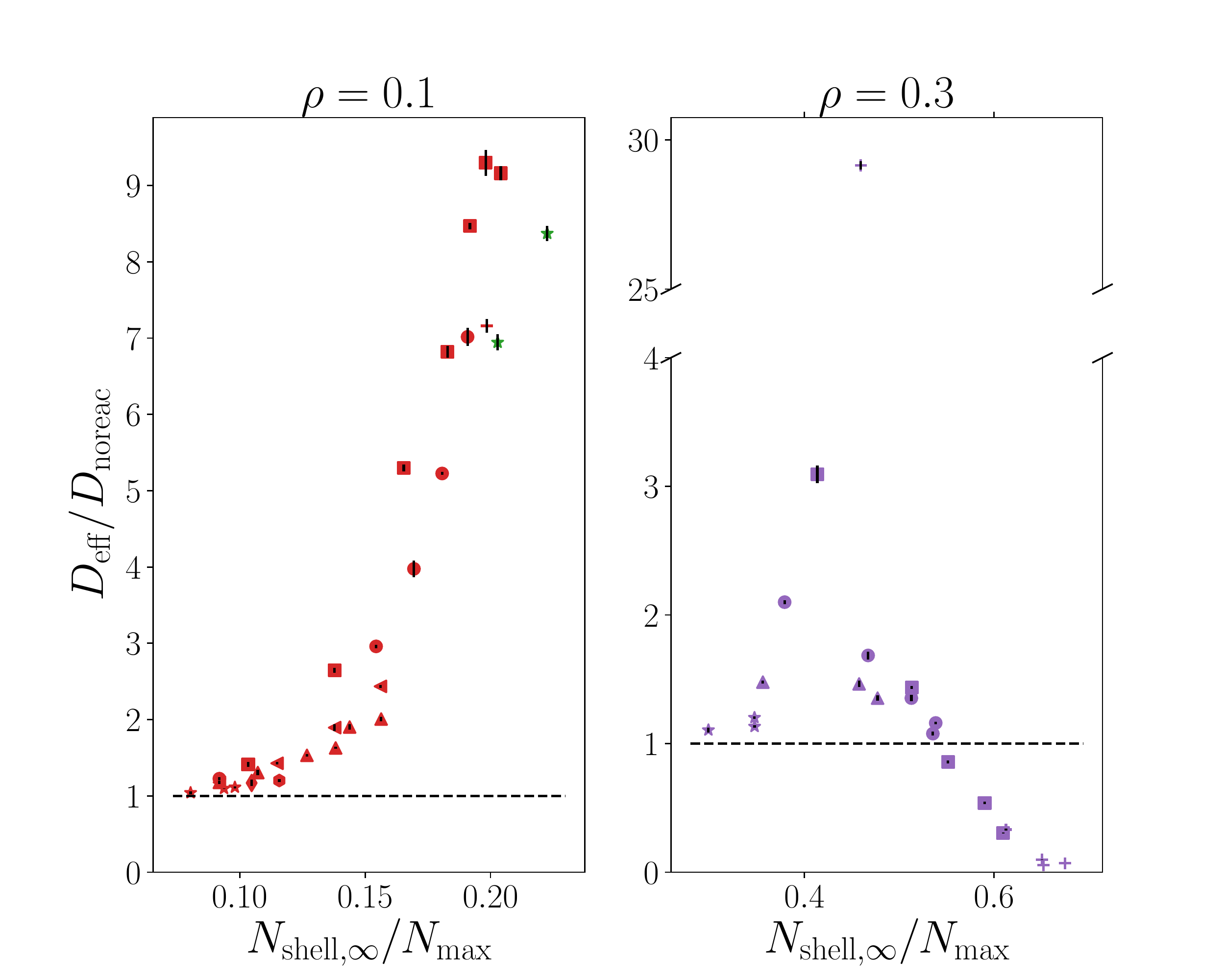}
\caption{Long-time diffusion coefficient of the colloid $D_{\mathrm{eff}}$ divided by the value without reaction $D_{\mathrm{noreac}}$ as a function of the filling fraction of the reaction area, measured by the stationary number of particles in the reaction area normalized by the maximum number of particles,  $N_{\mathrm{shell},\infty}/N_{\mathrm{max}}$. Left: results obtained with a solute density $\rho=0.1$ and right: solute density $\rho=0.3$.  All data, corresponding to different sets of parameters $\varepsilon$, and $r_{\rm cut}$, are collected here.  {Each symbol corresponds to a value of the Lennard-Jones parameter $\varepsilon$: $\star$: $1$; $\Diamond$: $1.5$; $\varhexagon$: $1.75$; $\bigtriangleup$: $2$;  $\triangleleft$: $2.25$; $\Circle$: $2.5$; $\square$: $3$; $+$: $5$}. The green stars correspond to the system with $\varepsilon=3$ and  $\tau_{\mathrm{BA}}=10$.  }
\label{D_func_N}
\end{figure} 
Figure~\ref{D_func_N} shows the long-time diffusion coefficient of the colloid divided by its value in the absence of reaction  as a function of $N_{\mathrm{shell},\infty}/N_{\mathrm{max}}$. The dashed line represents the threshold ($D_{\mathrm{eff}}/D_{\mathrm{noreac}}=1$) above which the colloid is active. All the data obtained for different values of $r_{\rm cut}$, $\varepsilon$ and $\rho$ are shown here. 
 Interestingly, it appears that using the filling fraction of the reaction area at steady state, measured here by the ratio $N_{\mathrm{shell},\infty}/N_{\mathrm{max}}$, we can collapse all the data on a single line, at a given solute density. Even the values of the effective diffusion coefficient obtained with a slower kinetics of the reaction $\B \to \A$, represented as green stars, collapse on the other data at $\rho=0.1$. Details on the influence of this kinetic parameter are given in Section \ref{section-TBA}. 

This representation underlines that the filling fraction is a key parameter that controls whether there is activity and  its intensity. The influence of the filling fraction of the reaction area is subtle, and we can distinguish three situations. As we have previously stated, the propulsion is linked to the polarity vector $\pp$, which is correlated with the force exerted by the solute particles on the colloid, and to the persistence time of its orientation $\tau_p$. At small filling fractions, solute density fluctuations in the vicinity of the colloid are very small, the polarization vector has a small amplitude and is characterized by a small persistence time $\tau_p$, thus the relative increase of the diffusion coefficient is limited. A second regime is observed at intermediate values of the filling fraction, where propulsion appears since there are enough particles in the vicinity of the colloid to form persistent droplets. This results in a highly enhanced diffusion. In this regime, both $\tau_p$ and $D_{\mathrm{eff}}/D_{\text{noreac}}$ are an increasing function of the filling fraction. At relatively low bath concentration ($\rho = 0.1$, Figure~\ref{D_func_N}-left), only these two regimes are observed. At a larger bath concentration ($\rho = 0.3$, Figure~\ref{D_func_N}-right), a non monotonous behavior is interestingly observed, revealing a third regime: When the filling fraction reaches a critical value, the activity starts to decrease. This may be related to a decrease of the polarization of the liquid droplet. Indeed, as the droplet size increases, the liquid may more likely push on  opposite directions, thus averaging out the resulting force. For high filling fractions, activity is suppressed due to the formation of a dense crystal around the colloid. Surprisingly, there are still density fluctuations as the crystal is not centered on the colloid.
Large values of $\tau_p$ are obtained in this case, but the side of the crystal situated in the same direction as the polarization vector hinders the motion of the colloid and suppresses the activity. This confirms that the filling fraction is an important parameter in the propulsion. {A special case occurs for $\rho=0.3,\varepsilon=5$ and $r_{\text{cut}}=5.5$ for which
a huge propulsion is observed, with $D_{\mathrm{eff}}/D_{\text{noreac}}$ close to $30$. This is actually a particular combination of the parameters that leads to this situation.
The solute density is high, and there is a strong attraction between solute particles, but the reaction area is small, and only $1$ or $2$ solute particles can fit in the shell. This leads to a particular shape of the LJ fluid inside the reaction area: the fluid surrounds the colloid, but cannot not form a close shell because of the geometric constraints. Consequently, the colloid is propelled through the open part of the shell, which favors a strongly enhanced diffusion.}

\section{Robustness of the activity to generic changes in the geometry of the reactive zone}
\subsection{Effect of the size of the colloid}
\label{section-size}

\begin{figure}[!h]
\includegraphics[width=\columnwidth]{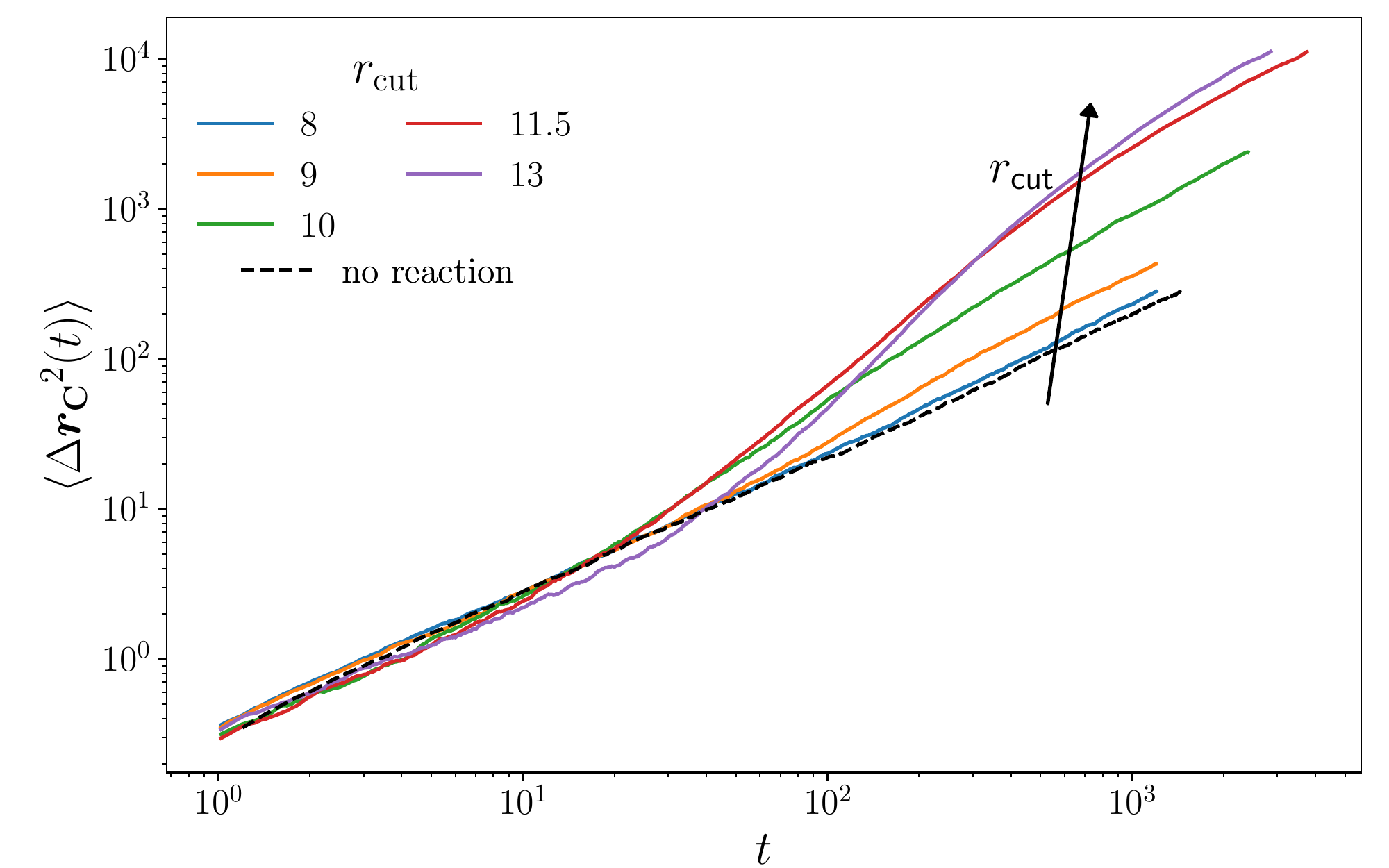}

\caption{Mean squared displacements of the colloid  as a function of time in a log-log scale for different value of $r_{\mathrm{cut}}$ at $\rho=0.1$ and $\varepsilon=3$. Here $\sigma_{\mathrm{C}}=10$. Note that the MSD has been computed taking an initial time corresponding to the steady state of the system. Here the plot shows times for $t>1$ to zoom on the ballistic and enhanced diffusion parts.}
\label{MSD-sigma10}
\end{figure}

In this part, we investigate the influence of the size of the colloid. Increasing its radius enhances the size asymmetry between colloid and solute particles. This results in a change of the curvature of the reactive zone that may affect the mechanism of droplet formation. The colloid diameter is increased by a factor $2$ ($\sigma_{\C}=10$ instead of $\sigma_{\C}=5$ previously). Our specific goal here is to check whether activity is still present, and whether the dependence of activity on the local solute density follows the same trend for both colloid sizes. Therefore, we chose parameters of the LJ fluid for which the smaller colloid was found to be active. For each system presented in this section, the average solute density is $\rho=0.1$, and we take $\varepsilon=3$ as the diffusion was strongly enhanced for these parameters for a smaller colloid.  The values of $r_{\mathrm{cut}}$ are larger than in the previous section to account for the larger size of the colloid. The values of $r_{\mathrm{cut}}$ vary between $8$ and $11.5$, these values correspond to reaction areas of about the same thickness as those obtained with $r_{\rm cut}=5.5$ and $r_{\rm cut}=9$ for a colloid twice smaller.

\begin{figure}
\includegraphics[width=\columnwidth]{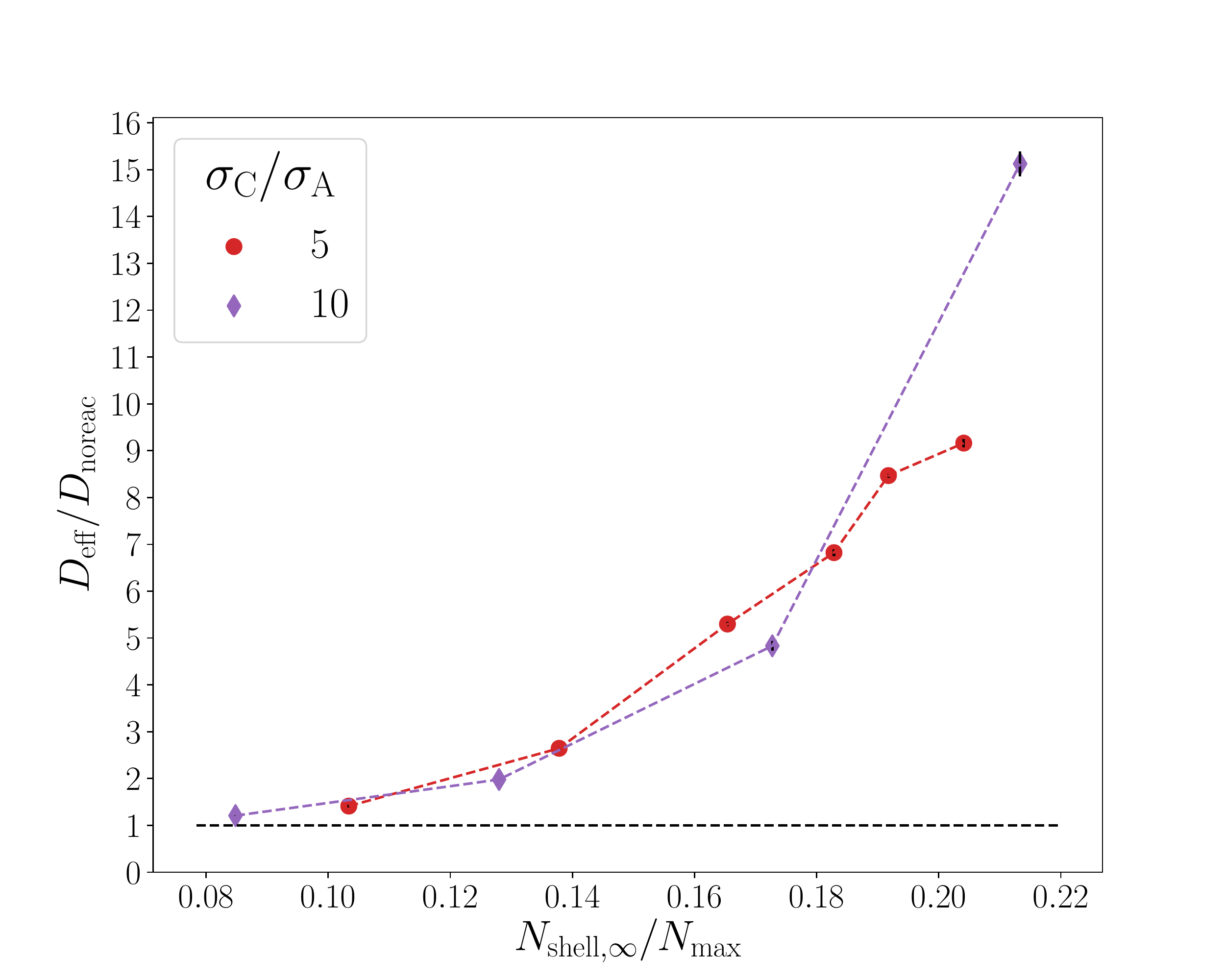}
\caption{Long-time diffusion coefficient of the colloid divided by its value without reaction, $D_{\rm eff}/D_{\mathrm{noreac}}$,  as a function of the stationary filling fraction of the reaction area, measured here by  $N_{\mathrm{shell},\infty}/N_{\rm max}$. Results obtained for both ratios $\sigma_{\mathrm{C}}/\sigma_{\rm A}$ are displayed, for $\rho=0.1$ and $\varepsilon=3$.}
\label{Deff-Nshell-bothsigma}
\end{figure}

We display on Fig. \ref{MSD-sigma10} the mean squared displacements of the colloid as a function of time in a log-log scale. Note that the initial time of this plot  has been actually rescaled as stated before,  and does correspond to the steady state. As already observed for $\sigma_{\C}=5$, the MSD has an intermediate ballistic regime, and the colloid displays an enhanced diffusion with a ratio $D_{\mathrm{eff}}/D_{\mathrm{noreac}}=15$ at long times. The intermediate ballistic regime is particularly visible for the highest reaction area (red plot). We give on Fig. \ref{Deff-Nshell-bothsigma} the values of the long-time diffusion coefficient of the colloid as a function of the filling fraction of the reaction area at steady state $N_{\mathrm{shell},\infty}/N_{\rm max}$: The results obtained with both ratios of $\sigma_{\C}/\sigma_{\A}$ coincide very well. Again, we see that enhanced diffusion is mainly controlled by the filling fraction of the reaction area.  
All in all, we observe no significant changes while increasing the ratio $\sigma_{\C}/\sigma_{\A}$ by a factor two: The mechanism that causes an enhanced diffusion seems to be a robust feature of our model system.

\subsection{Influence of the kinetics of the backward reaction $\B \to \A$}
\label{section-TBA}

Another important parameter of our model is $\tau_{\mathrm{BA}}$, which controls the kinetics of the reaction $\B\to \A$. In particular, one may wonder whether the activity sill holds when this reaction is slower, which is expected to create a cloud of B particles behind the colloid as it moves. To investigate this, we consider the case 
$\rho=0.1$, $\varepsilon=3$ and we change the reverse reaction rate in order to make the reverse reaction $100$ times slower: $\tau_{\mathrm{BA}}=10$ instead of $\tau_{\mathrm{BA}}=0.1$. We restrict our study to two reaction areas: 
$r_{\mathrm{cut}}=7.5$ and $r_{\mathrm{cut}}=10.5$.

\begin{figure}%[h!]
\center
\includegraphics[width=\columnwidth]{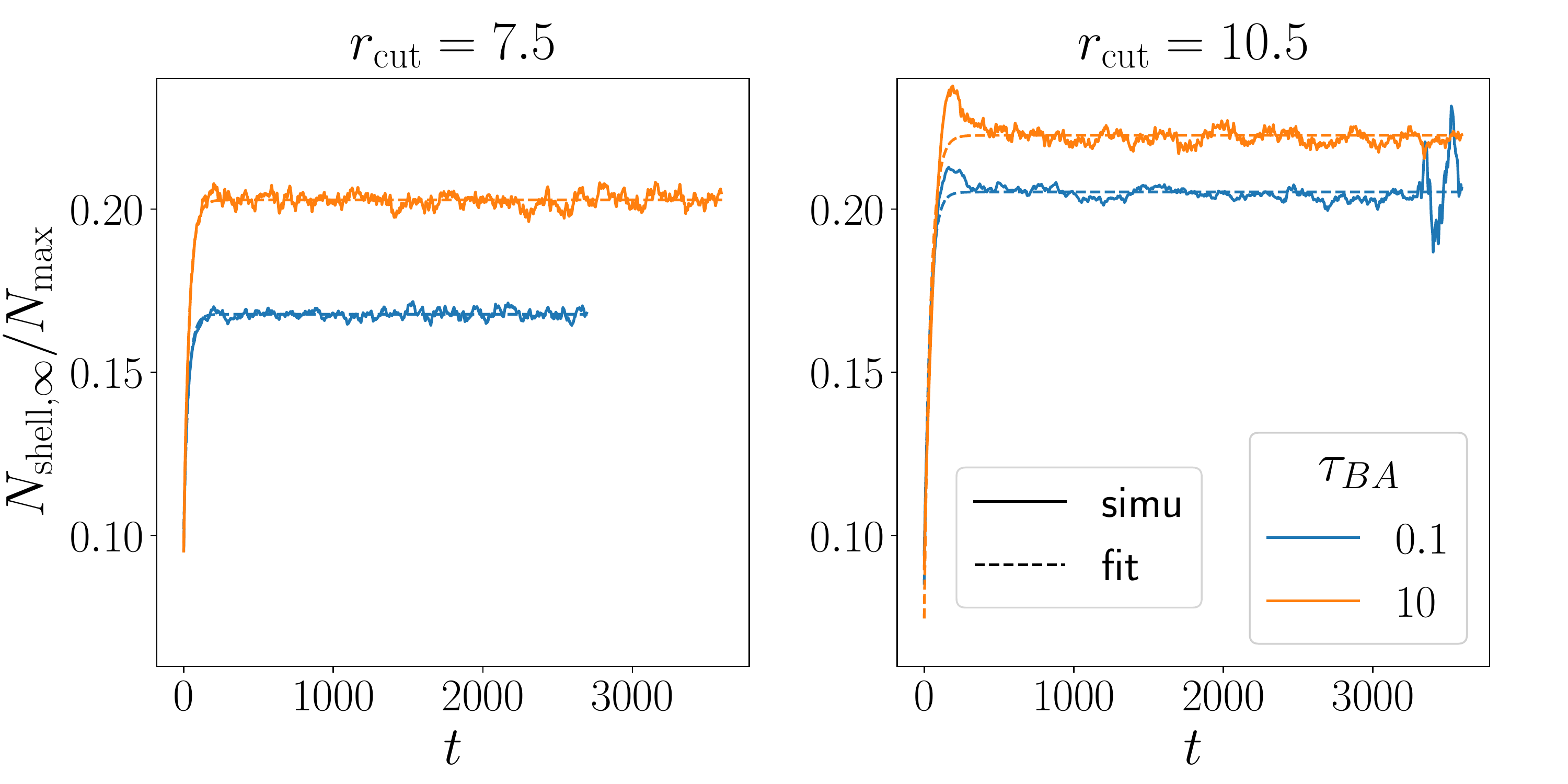}
\caption{Stationary number of particles in the reaction area normalized by the maximum value, $N_{{\rm shell},\infty}/N_{\rm max}$ as a function of time for different values of $\tau_{\mathrm{BA}}$. The dashed line is the fit using equation~\eqref{eq_n_shell}. The parameters of the model are: $\rho=0.1$, $\varepsilon=3$, left : $r_{\mathrm{cut}}=7.5$, right : $r_{\mathrm{cut}}=10.5$.}
\label{n_shell_tau_BA}
\end{figure}

\begin{figure}%[h!]
\includegraphics[width=\columnwidth]{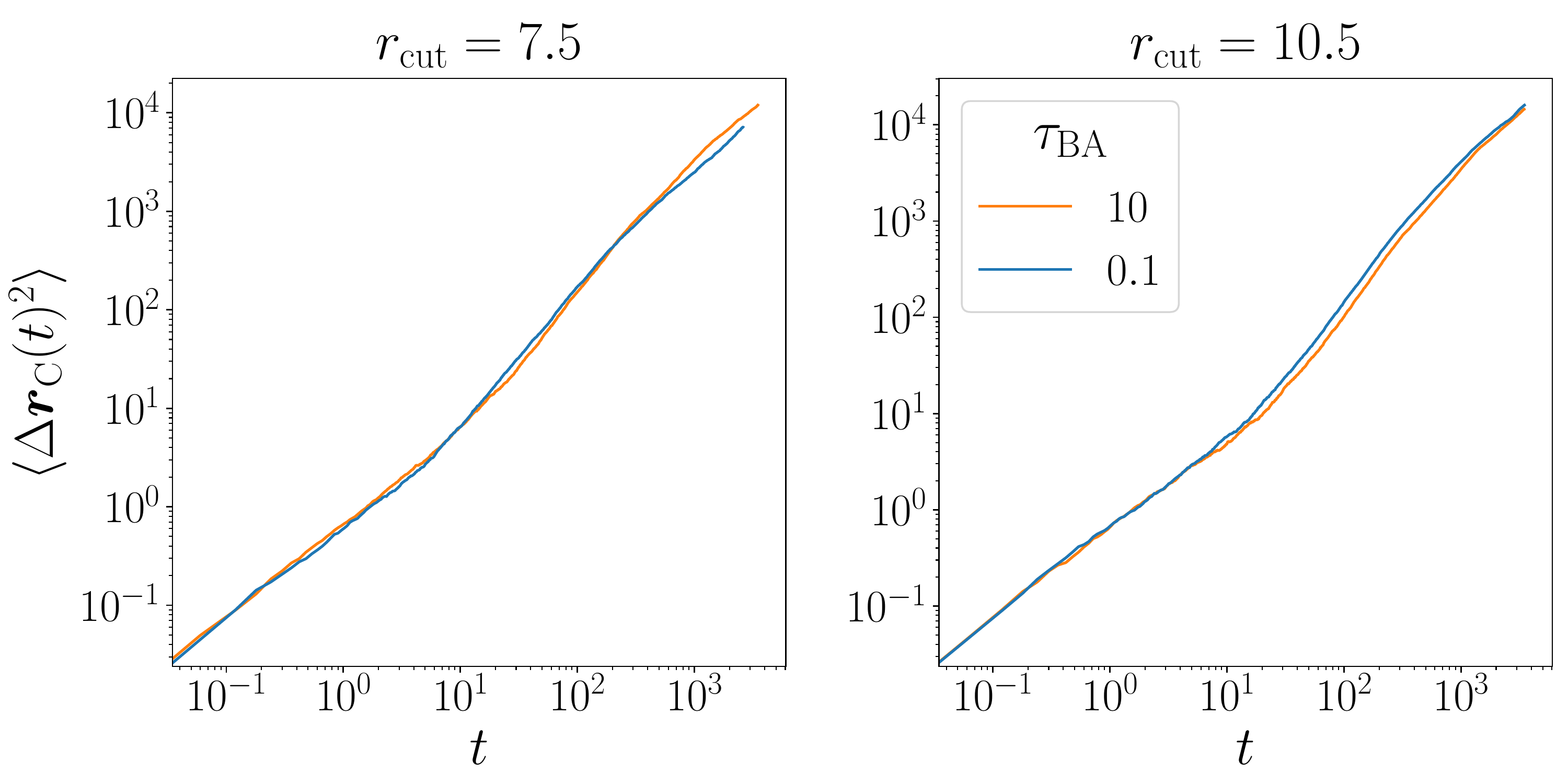}
\caption{Mean squared displacements of the colloid as a function of time, in a log-log scale for the systems with  $\rho=0.1$, $\varepsilon=3$, $\sigma_{\rm C}=5$ and two different values of the characteristic time $\tau_{\rm BA}$ of the reverse reaction $\B \to \A$. Left:  $r_{\mathrm{cut}}=7.5$. Right: $r_{\mathrm{cut}}=10.5$. Note that the initial time here corresponds to the steady state of the system. }
\label{msd_tauBA}
\end{figure}

\begin{figure}%[h!]
\includegraphics[width=\columnwidth]{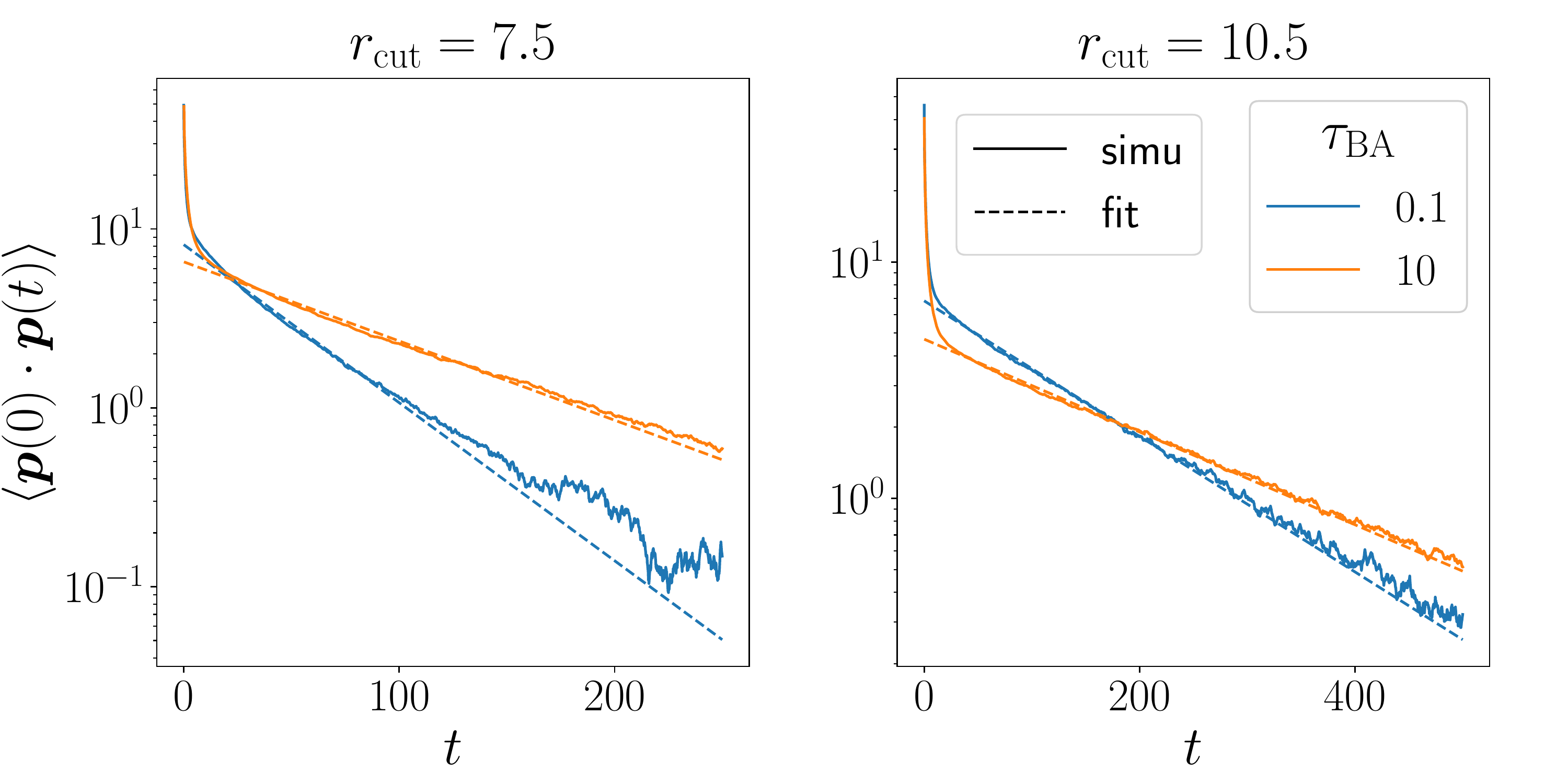}
\caption{Auto-correlation function of the polarization vector $\boldsymbol{p}$ defined by Eq. \ref{def-p} as a function of time in a log-log scale,  for the systems with  $\rho=0.1$, $\varepsilon=3$, $\sigma_{\rm C}=5$ and two different values of the characteristic time $\tau_{\rm BA}$ of the reverse reaction $\B \to \A$. Left:  $r_{\mathrm{cut}}=7.5$. Right: $r_{\mathrm{cut}}=10.5$. Note that the initial time here corresponds to the steady state of the system. }
\label{taup_tauBA}
\end{figure}

First, we monitor the amount of particles in the reaction area as a function of time. Results are displayed on Fig.~\ref{n_shell_tau_BA}. For both values of $r_{\rm cut}$, we observe that the filling fraction of the reaction area increases when the kinetics of the reverse reaction is slown down. Exponential fits agree well with simulation data, and the characteristic time needed to reach the steady state, $\tau_{\text{N}}$, is slightly increased when the reverse reaction is slown down (see values in Table~\ref{table-tau-BA}). The system under these conditions is in a range where $D_{\rm eff}/D_{\mathrm{noreac}}$ is an increasing function of the filling fraction. The MSD indeed confirms that activity is still present, as it can be seen on Fig. \ref{msd_tauBA}. The colloid motion again displays an intermediate ballistic regime. However, the effective diffusion coefficient is not significantly increased. Fig. \ref{taup_tauBA} shows the auto-correlation function of the polarization vector $\pp$ for the systems investigated here (in a semi-log scale). In every case, this function becomes exponential at long times, and we can calculate the characteristic persistence time $\tau_p$ from the slope. It clearly appears that this correlation time increases for both values of $r_{\rm cut}$. This can be intuitively understood: When the reverse reaction is slowed, the total number of $\B$ solute particles in the system is higher than $N_{{\rm shell},\infty}$ and the LJ fluid expands over the edge of the reaction area. This yields the formation of a larger droplet of B particles in the vicinity of the colloid, which acts as an effective $r_{\mathrm{cut}}$ higher than the real one. If the reaction area becomes too large, the droplet can come apart from the colloid surface, decreasing the force induced by the polarization vector $\pp$, and thus the effective diffusion coefficient is not significantly increased.

\begin{table}
\begin{center}
\begin{tabular}{@{}l|  c c|c c|}
\cline{2-5}
\rule{0mm}{5mm}
 &  \multicolumn{2}{c|}{$\tau_{\rm BA}=0.1$} &  \multicolumn{2}{c|}{$\tau_{\rm BA}=10$} \\ 
 \rule{0mm}{5mm}
 & $\tau_{\mathrm{N}}$& $N_{\rm shell,\infty}$ & $\tau_{\mathrm{N}}$& $N_{\rm shell,\infty}$ \\ 
\hline 
\vline\rule{0pt}{5mm}$r_{\mathrm{cut}}=7.5$ & $31$ & $30$ & $35$ & $37$ \\ 
\vline\rule{0pt}{5mm}$r_{\mathrm{cut}}=10.5$ &$36$ & $77$ & $41$ & $84$  \\
\hline 
\end{tabular}
\end{center}
 \caption{Fitted values of Eq.~\eqref{eq_n_shell} $\tau_{\mathrm{N}}$, and $N_{\rm shell,\infty}$, for systems with $\rho=0.1$, $\varepsilon=3$, $\sigma_{\C}=5$ and different reaction rates $\tau_{\mathrm{BA}}$.}
\label{table-tau-BA}
\end{table} 

\subsection{Propulsion in three dimensions}
\label{section-3D}

\begin{figure}%[h!]
\includegraphics[width=\columnwidth]{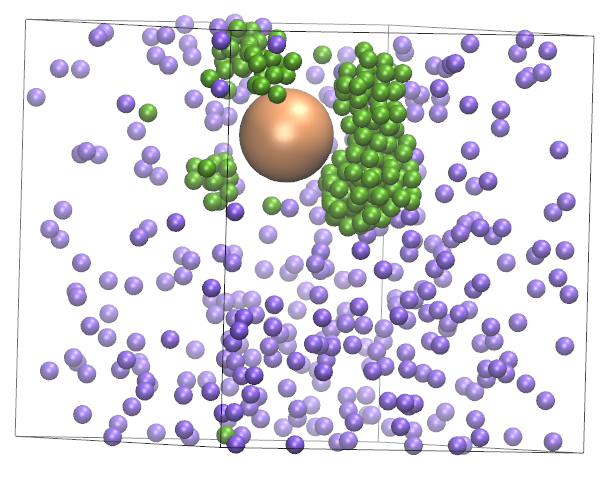}
\caption{Snapshot of the system studied in three dimensions, at steady state. The density of solute particles is $\rho=0.05$, the intensity of the LJ attraction is $\varepsilon=3$, and the size of the reaction area is $r_{\rm cut}=7.5$.}
\label{snapshot_3D}
\end{figure}

\begin{figure}%[h!]
\includegraphics[width=\columnwidth]{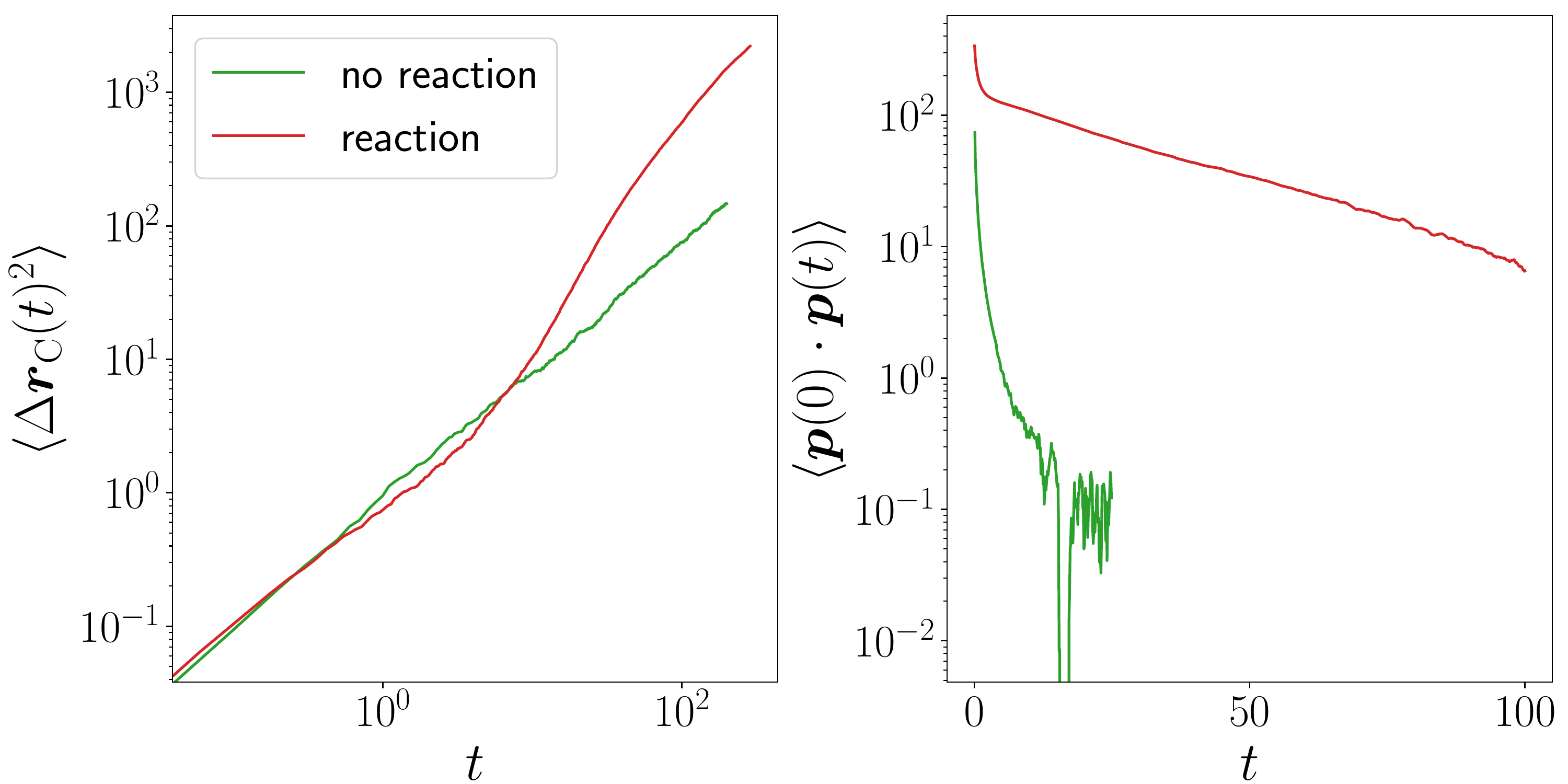}
\caption{Dynamic quantities obtained at steady state for a 3D system with the solute density $\rho=0.05$, the LJ attraction $\varepsilon=3$, and the reaction area of radius $r_{\rm cut}=7.5$. The case with reaction, that displays activity (in red) is compared to the case without reaction (in green). Left: Mean squared displacement of the colloid as a function of time in a log-log scale. Right: Auto-correlation function of the polarization vector $\pp$ as a function of time for the system with reaction.} \label{MSD_3D}
\end{figure}

So far, all the presented results concern 2D simulations. As the mechanism of droplet formation and the persistence time of the droplet may depend on the dimension of the system, 
we also performed simulations for the same model in three dimensions. The simulation box is cubic with $l_{\rm box}=22$. We chose a solute density $\rho=0.05$, an interaction parameter $\varepsilon=3$ and a reaction radius $r_{\rm cut}=7.5$. A snapshot of the system at steady state is shown on Fig.~\ref{snapshot_3D}. Qualitatively, we observe on the figure the appearance of droplets of B particles around the colloid, as was obtained in 2D. The mean squared displacement of the colloid  as a function of time at steady state is given on Fig.~\ref{MSD_3D}-left (again, where the time was rescaled as previously described, and the plot begins at steady state). We observe a  behavior similar to 2D systems : The MSD has a ballistic motion at intermediate times and becomes linear in time at long times.  Importantly, we obtain again an enhanced diffusion at long time with $D_{\rm eff}/D_{\mathrm{noreac}} \sim 11$ for the parameters used here. The auto-correlation function of the polarization vector $\pp$ is shown on Fig.~\ref{MSD_3D}-right in a semi-log scale. It is exponential at long times, and characterized by a  persistence  time $\tau_p$ of the same order of magnitude as in 2D \cite{Decayeux2021a}. It appears thus that all the essential features of the 2D propulsion mechanism still hold at 3D: Droplets form in the reaction area, which push the colloid with a persistent orientation.

\section{Conclusion}
We propose a model in which activity arises from density fluctuations in the vicinity of an isotropic colloid. In this article we clarify the influence of the relevant parameters for the emergence of self-propulsion. We show the results of simulations performed for a wide range of parameters. We highlight that the filling fraction of the reaction area $N_{\mathrm{shell},\infty}/N_{\rm max}$ is a key parameter to determine whether there is activity and its intensity. We find that we can collapse all the data using this observable. We show that our mechanism is robust as activity still occurs while increasing by two the relative size $\sigma_{\C}/\sigma_{\A}$, and slowing the reverse reaction down by a factor 100. The mechanism also still holds in three dimensions. In a future work we will use this mechanism and study the collective dynamics of several active colloids along with the structural properties \cite{Bialke2015}. {In particular, this will be the opportunity to discuss the links between our model and usual models for active particles, such active Brownian particles \cite{Bechinger2016}.}

\section*{Author contribution statement}
JD made the investigation. JD did the data curation. All authors contributed to the methodology, the formal analysis, and writing.  

\section*{Data Availability Statement}
The datasets generated during and/or analysed during the current study are available from the corresponding author on reasonable request.

\bibliographystyle{apsrev4-1}
\bibliography{draft_catalytic_colloid}

\end{document}